\documentclass[onecolumn,preprintnumbers,nofootinbib]{revtex4}
\usepackage{sidecap}
\usepackage{wrapfig}

\usepackage{graphicx}
\usepackage{amssymb}
\usepackage{color}
\usepackage{enumerate}
\usepackage{amsmath}
\usepackage{slashed}
\usepackage{comment}
\def\beq{\begin{equation}}
\def\eeq{\end{equation}}
\def\beqn{\begin{eqnarray}}
\def\eeqn{\end{eqnarray}}

\def\e{\epsilon}

\renewcommand{\texttt}{{}}
\newcommand{\be}{\begin{eqnarray}}
\newcommand{\ee}{\end{eqnarray}}

\begin{document}

\title{
\bf  {Super-renormalizable or Finite Lee\,-Wick Quantum Gravity 
}}

\author{Leonardo Modesto} 
\email{lmodesto@fudan.edu.cn,  lmodesto1905@icloud.com}

%\ead{lmodesto@fudan.edu.cn}

%\author{Ilya L. Shapiro$^*$}
%\ead{ilyashapiro2003@yahoo.com.br}

\affiliation{Center for Field Theory and Particle Physics and Department of Physics, \\
Fudan University, 200433 Shanghai, China   \\
%{\rm email: lmodesto@fudan.edu.cn} %, rachwal@fudan.edu.cn}
}

%\address{
%{%$^{*}$\small Departamento de Fisica - ICE, Universidade Federal de Juiz de Fora 33036-330, Juiz de Fora - MG, Brazil
%}
%}

%\date{\small\today}

\begin{abstract} \noindent
We propose a class of multidimensional higher derivative theories of gravity 
without extra real degrees of freedom besides the graviton field. 
%The most general action is
%marked by three local polynomial of the d'Alembertian operator that averts 
%extra real degrees of freedom, and improve the high energy
%behaviour of the loop amplitudes. 
The propagator shows up the usual real graviton pole in $k^2=0$ and extra 
complex conjugates poles that do not contribute to
the absorptive part of the physical scattering amplitudes.
Indeed, they may consistently be excluded from the asymptotic  observable states of the theory 
making use of the Lee-Wick 
and Cutkoski, Landshoff, Olive and Polkinghorne 
prescription for the construction of a unitary S-matrix. 
Therefore, the spectrum consists on the graviton and short lived elementary unstable particles 
that we named ``anti-gravitons" because of their repulsive 
contribution to the gravitational potential at short distance.  
However, another interpretation of the complex conjugate pairs is proposed  
based on the 
Calmet's suggestion, i.e. they could be understood as black hole precursors long established in the classical theory. 
Since the theory is CPT invariant, the complex conjugate of the micro black hole precursor 
has received as a white hole precursor consistently with the t'Hooft complementary principle. 
It is proved that the quantum theory is super-renormalizable in even dimension, i.e.
only a finite number of divergent diagrams survive, and finite in odd dimension.  
%%
%Furthermore, in odd dimension (for theories convergent from two loops upward) there
%are no counter terms we can construct out of the Riemann tensor  and the theory turns out to be ``finite". 
%Finally, 
Furthermore, 
turning on a local potential of the Riemann tensor we can make the theory finite in any dimension. 
The singularity-free Newtonian gravitational potential is explicitly computed for a range of higher derivative theories. 
%We end up with a proposal for 
Finally, we propose a new super-reneromalizable or finite Lee-Wick standard model of particle 
physics.

\end{abstract}

\maketitle

\vspace{-1.2cm}

%\tableofcontents

\section{Introduction}

We propose a ``local" multidimensional gravitational theory 
compatible with renormalizability at perturbative level in addition to 
Lee-Wick \cite{LW}
and Cutkoski, Landshoff, Olive and Polkinghorne 
unitarity \cite{Cutk} (CLOP). %, as a forthcoming of the %in the ``quantum field theory framework". 
This work is a generalization of 
the theory recently proposed in \cite{SM, ComplexGhostsIS}. 
In the last four years % some recent papers \cite{modesto,BM,Bis2, M2, M3, M4}, %but in particular in \cite{modesto} 
a weakly nonlocal action principle for gravity has been extensively studied 
to make up for the shortcomings of the quantization of the Einstein-Hilbert action 
\cite{modesto,%BM,Bis2, 
M2, M3, M4, CalcagniMtheory}.
%Such path to quantum gravity was initially inspired by Cornish and Moffat papers \cite{Moffat} %,corni1,corni2,Moffat2,corni3,Moffat4}
%as well as more recently 
%and afterwords motivated by Tomboulis studies on weakly nonlocal super-renormalizable gauge and gravitational theories \cite{Tombo}. % and quantum gravity \cite{Tombo70, Tombo80, TomboAnto}.
Research records show that Krasnikov in 1988 and Kuz'min in 1989 proposed a similar theory  \cite{Krasnikov},
following Efimov's %studies in nonlocal  interacting quantum field theory 
\cite{efimov}. %,E2,E3,E4,E5}. 
Afterwords Tomboulis extended to gauge interactions the Kuz'min ideas and in $1996$ proposed a class of weakly nonlocal super-renormalizable gauge and gravitational theories \cite{Tombo,Tombo2,Tombo3}. 
%See also more recent papers by Tomboulis \cite{Tombo2,Tombo3}. % and more recently by other authors 
You may also refer to 
\cite{calcagniNL,Maggiore} about other excellent contributions in nonlocal theories.
Recently in \cite{LeslawModestoFinite} has been definitely proved that the theory is actually finite in 
any dimension wether a local potential of the Riemann tensor is added. 
In \cite{PivaFGT} has been proposed and extensively studied a finite generalization of the nonlocal theory for gauge interactions proposed for the first time by Tomboublis \cite{Tombo}.  
However, the price to pay is that the classical action is weakly nonlocal, although the 
asymptotic polynomial behaviour makes the theory very similar to any local higher derivative theory for all that concerns the divergent contributions to the quantum effective action. 

In this paper we want to expand and specialize the seminal paper \cite{shapiro3} about a general 
local super-renormalizable gravitational theory capitalizing what we learned in quasi-polynomial or weakly nonlocal theories. 
% about unitarity. 
Actually, many results can be exported directly to the theory here proposed making %when it is made 
a proper replacement of the {\em nonlocal form factor}  
in \cite{LeslawModestoFinite} with the {\em local form factor} that we are going to properly define later in this paper.

The theory here proposed fulfills a synthesis of minimal requirements: 
(i) Einstein-Hilbert action should be a good approximation of the theory at a much smaller energy scale than the Planck mass;  %the action has to be well approximated by the, % Einstein-Hilbert action,
%(iii) 
%\item the spacetime dimension has to decrease with the energy in order to have %, at least in principle,
%a complete quantum theory in the ultraviolet regime; %this hypothesis is 
%(iv) 
%\item 
(ii) 
the theory has to be %perturbatively 
super-renormalizable %or renormalizable 
or finite at quantum level; 
%(this hypothesis is strongly related to the previous one);
%(v) 
%\item 
(iii) the theory has to be unitary, with no other real 
%poles in the propagator then the graviton 
%degree of freedom 
poles in the propagator in addition to the graviton; % in the propagator, 
if we require other poles nether real nor complex, then the theory results non-polynomial or weakly nonlocal. 
%(vi) 
%\item spacetime is a single continuum of space and time and in particular the 
%Lorentz invariance is not violated. %broken. 
%\end{enumerate}
%}
%The last requirement is supported by observations.
The outcome of previous studies is a nonlocal classical theory of gravity
perturbatively super-renormalizabile at quantum level.  
On the footprint of the nonlocal action we propose here a ``local" theory that holds the
same properties, but showing up extra complex conjugate poles besides the graviton.

Studies of higher derivative theories date back to quadratic gravity 
%A first local extension of Einstein gravity compatible with quantum mechanics was 
proposed in 1977 by Stelle \cite{Stelle}. 
This theory is renormalizable and asymptotically free, but unfortunately it violates unitarity 
showing up a real ghost state in the spectrum. 
In this paper we go behind the Stelle's action introducing a finite number of extra higher derivative operators to make
the theory even more convergent: super-renormalizable or finite. However, we do not blandly introduce all the possible operators 
to a fix order in the number of derivatives of the metric tensor. We actually consider a class of local
theories that avoid extra real poles in the propagator. 
Looking at the above list of requirements (i)-(iii), 
the news respect to the previous work on non-polynomial theories sits in the third point. 
We indeed do not exclude the possibility 
of complex conjugate mass poles, which do not prevent 
us from constructing a unitary local theory of gravity 
in the Lee-Wick formalism \cite{LW}.
Lee and Wick argued that, as long as all ghost degrees of freedom in the interacting theory have complex energies, one obtain a unitary theory by constraining the physical subspace to be exactly the one for
the states that have real energy.
In gravity we end up with a classical theory with an extended spectrum in which the graviton is free to propagate on long distances while a bunch of
other virtual elementary particles can only intrinsically live for a short amount of time \cite{Veltman}.
It is well known, that in quantum electro-dynamics 
a photon can get converted into $e^+$$e^-$ pairs, or more complicated channels, only when it interacts with matter, but when radiated into a ``perfect vacuum" it will travel on indefinitely distances as a stable particle. 
In field theory this is described by a gauge independent pole at $k^2 =0$ in the transverse photon propagator, which fixes the photon free field equations to $\Box A_{\mu}=0$. 
By the contrast, we here have a finite number of short lived particles (named ``anti-gravitons")
% or ``angravitons")
that rapidly convert themselves in gravitons. The dispersion relation for these particles must show off a finite lifetime
through gauge independent complex poles in the propagator, and the free equations of motion are
$[\Box + (A+iB)] \phi = 0$ (where $A,B \in \mathbb{R}$.)
In particle physics a Lee-Wick extension of the standard model has been proposed 
to avoid quadratic divergences in the Higgs mass and hence no hierarchy puzzle \cite{Wise2}.
In this theory, and generalizations, the classical action has a real ghost pole that, at one loop, is shifted out the real axes 
into a complex ghost pair. 
In gravity a similar Lee-Wick unitarization of the Stelle's theory \cite{Stelle} was evoked in 
\cite{Tombo70, Tombo80, TomboAnto} to remove the real ghost from the asymptotically free quadratic gravity \cite{AF}. 
Indeed, at one loop the real ghost pole splits in a pair of complex conjugate poles. 
In this paper we go beyond four derivatives and following
the seminal papers \cite{Jansen, Yamamoto1} 
%, Veltman}\footnote{In \cite{Veltman} 
%
%the author has shown for the case of a super-renormalizable theory with an unstable particle that the 
%$S$-matrix is unitary 
%(in non-perturbative QFT) on the Hilbert space spanned by only the stable particles.}
%
%
we propose a theory in which a finite number of complex conjugate poles,
or unstable particles, are already present 
in the classical action \cite{Liu}. %at classical level. 
Let us to give here a taste of the theory\footnote{{Definitions and notations}. 
The definitions used in this paper are: the metric tensor $g_{\mu \nu}$ has 
signature $( + - \dots -)\,$; the curvature tensor  
$R^{\mu}_{\nu \sigma \rho } = - \partial_{\rho} \Gamma^{\mu}_{\nu \sigma} + \dots$, 
the Ricci tensor $R_{\mu \nu} = R^{\rho}_{\mu\rho\nu}$, and the curvature scalar 
$R = g^{\mu\nu} R_{\mu\nu }$.  We also use the notation $\mathcal{R}$ for the Riemann tensor when the
indexes are suppressed.}, %the quite general structure of the theory,
\be 
 %\hspace{-0.6cm}
S_{\rm SR}  = - \int \! d^D x \sqrt{|g|} \, 2 \kappa_D^{-2} \Big[ {\bf R} - 2 \Lambda_{\rm cc} % {\bf \bar{\lambda}}
 + {\bf G  \gamma (\Box) 
 Ric } + {\bf V} \Big]   .
 \label{Daction}
 \ee 
where ${\bf \gamma (\Box)}$ is a polynomial (of the d'Alembertian operator $\Box$) constructed so as to avoid 
extra real poles in the propagator besides the graviton, ${\bf G}$ is the Einstein tensor, ${\bf Ric}$ 
is the Ricci tensor, and ${\bf V}$ is a potential at least cubic in the curvature tensor. 

At classical level the solutions are stable when Lee and Wick %proposed to remove the instability by imposing 
appropriate boundary conditions are imposed \cite{LW, Cutk}.
More recently a mathematically well defined prescription 
has been defined in \cite{Barnaby}. However, microcausality is violated.

%Actually in \cite{Yama} the author proves that the scattering matrix is unitary in the 
%subspace of real massive particles, without to refer to the Lee-Wick 
%does not refer to 
 %
%The interpretation here evoked is inspired by .. 
%

%\paragraph*
\section{The theory} % ---}\
\label{TheTheory}
%quantum gravity}

%%%%%%%%%

%%%%%%%%

The class of theories we are going to propose can be read out from the 
``non-polynomial" theories recently introduced and extensively studied in \cite{Tombo, modesto,LeslawModestoFinite}. 
We here focus on a general local action compatible with unitarity \cite{LW, Yamamoto1} 
and super-renormalizability or finiteness, 
%, focusing mainly on the quadratic part in the curvature, reads
%
%
\be
&& 
\mathcal{L}_{\rm g} = -  2 \kappa_{D}^{-2} \, \sqrt{g} 
\left[ {\bf R} 
+
{\bf R} \, 
 \gamma_0(\Box)%_{\Lambda})
 {\bf R} 
 + {\bf Ric} \, 
\gamma_2(\Box)%_{\Lambda})
 {\bf Ric} 
+ {\bf Riem}  \, 
\gamma_4(\Box)%_{\Lambda}) 
{\bf Riem} 
+ {\bf V} \, 
\right]  .
\label{gravityG}
\ee
We can rewrite the theory making use of a more compact notation introducing a tensorial form factor, namely
\be
&& \hspace{-1.3cm}
\mathcal{L}_{\rm g} = -  2 \kappa_{D}^{-2} \, \sqrt{g} 
\left( {\bf R} 
+ {\bf Riem}  \, %\cdot
{\bf \gamma}(\Box)%_{\Lambda}) 
{\bf Riem} 
+ {\bf V} 
\right)  \nonumber \\
&& \hspace{-0.8cm}
\equiv 
-  2 \kappa_{D}^{-2} \, \sqrt{g} 
\left(    R 
+ {R}_{\mu\nu\rho\sigma}  
\gamma(\Box)^{\mu \nu \rho \sigma}_{\alpha \beta \gamma \delta} \, 
{R}^{\alpha\beta\gamma\delta} 
+ {\bf V} 
\right)
\label{gravityG2} \\
&& \hspace{-1.3cm}
\equiv 
-  2 \kappa_{D}^{-2} \, \sqrt{g} 
\left\{    {\bf R} 
+ {R}_{\mu\nu\rho\sigma}  \, 
[  g^{\mu\rho} g^{\alpha\gamma} g^{\nu \sigma} g^{\beta \delta} \gamma_0(\Box)
+ g^{\mu\rho} g^{\alpha\gamma} g^{\nu\beta} g^{\sigma\delta} \gamma_2(\Box)  
+ g^{\mu\alpha} g^{\nu\beta} g^{\rho \gamma} g^{\sigma \delta} \gamma_4(\Box)  ]
{R}_{\alpha\beta\gamma\delta} 
+ {\bf V} 
\right\} .
\nonumber
\ee
The theory consists on a kinetic local operator quadratic in the curvature, three polynomials 
$\gamma_0(\Box)$, $\gamma_2(\Box)$, $\gamma_4(\Box)$, and a local potential ${\bf V}$ 
%at least cubic in the curvature.
%The local potential is 
made of the following three sets of operators, %reads 
%sector and a local curvature potential, namely 
%\begin{widetext}
\be
\hspace{0.1cm}
 {\bf V} = 
 \sum_{j=3}^{{\rm N}+2} \sum_{k=3}^{j} \sum_i c_{k,i}^{(j)} \left( \nabla^{2(j-k)} {\cal R}^k \right)_i
 \!
 +
  \!\!\!
 \sum_{j={\rm N}+3}^{\gamma+{\rm N}+1} \sum_{k=3}^{j} \sum_i d_{k,i}^{(j)} \left(\nabla^{2(j-k)} {\cal R}^k \right)_i
 + \!\!\!
 \sum_{k=3}^{\gamma +{\rm N}+2} \!\!
 \sum_i s_{k,i} \, \left(  \nabla^{2 (\gamma + {\rm N}+2 -k )} \, {\cal R}^k \right)_i ,  
 %\label{K0}
 \nonumber 
 \ee 
 where the third set of operators are called killers because they are crucial in making the theory finite in any 
 dimension.
 %In (\ref{K0}) 
 $\Lambda$ is an invariant mass scale and the indices,
 $c_{k,i}^{(j)}$, $d_{k,i}^{(j)}$, $s_{k,i}$ are running or not coupling constants, while the tensorial structure have been neglected. The last set of operators with front coefficients $s_{k,i}$ are technically called ``killers" and are crucial in making the theory finite. 
%
 % (in odd dimension we do not need to introduce any potential to make the theory finite), 
%The sum in (\ref{K0}) must include at least the minimal set of operators (with different tensorial structure), which we need to make the theory finite. 
The capital $\rm{N}$ is defined to be the following function of the spacetime dimension $D$: $2 \mathrm{N} + 4 = D$. 
Moreover, 
$\Box = g^{\mu\nu} \nabla_{\mu} \nabla_{\nu}$ is the covariant box operator, while $\gamma$ is a positive integer. The polynomials $\gamma_i(\Box)$ are:  
%$H(-\Box_{\Lambda})$ 
%will be shortly defined. 
\be
%&& 
%\hspace{-0.5cm}
\gamma_2(\Box) = - \!\! \sum_{i=1}^{{\rm{N} }+ \gamma+1} \!\! \frac{a_i}{\Lambda^2} z^{i-1} - 4 \gamma_4 \, , \quad % \nonumber \\&&  
\gamma_0(\Box) = \frac{ D-2}{4 (D-1)} \!\!  \sum_{i=1}^{{\rm{N} }+ \gamma+1} \!\! \frac{b_i}{\Lambda^2} z^{i-1} 
+ \frac{ D}{4 (D-1)} \!\!  \sum_{i=1}^{{\rm{N} }+ \gamma+1} \!\! \frac{a_i}{\Lambda^2} z^{i-1}
+ 4 \gamma_4  \, , 
\ee
where $z:= - \Box/\Lambda^2$ and we implicitly introduced the following two polynomials,
\be
 P(z) = 1 + \!\! \! \sum_{i=1}^{{\rm{N} }+ \gamma+1} \!\! a_i z^i \, , \quad  %\nonumber \\
Q(z) = 1 + \!\! \! \sum_{i=1}^{{\rm{N} }+ \gamma+1} \!\! b_i z^i . 
\label{PolyG}
\ee
The reason of this particular choice of the polynomials $\gamma_i(\Box)$ will be clear in the next subsection
when we explicitly evaluate the propagator for the theory (\ref{gravityG}). 
\subsection{Propagator}
Splitting the spacetime metric $g_{\mu\nu}$ 
in the flat Minkowski background and the fluctuation $h_{\mu \nu}$ 
defined by $g_{\mu \nu} =  \eta_{\mu \nu} + \kappa_D \, h_{\mu \nu}$,
we can expand the action (\ref{gravityG}) to the second order in $h_{\mu \nu}$.
The outcome of this expansion together with the usual harmonic gauge fixing term reads \cite{HigherDG}
%\be
$\mathcal{L}_{\rm lin} + \mathcal{L}_{\rm GF} 
= 1/2h^{\mu \nu} \mathcal{O}_{\mu \nu, \rho \sigma} \, h^{\rho \sigma}$,
%\label{O}
%\ee
where the operator %inverse of the propagator 
$\mathcal{O}$ is made of two terms, one coming from the linearization of (\ref{gravityG})
and the other from the following gauge-fixing term,
$\mathcal{L}_{\rm GF}  = \xi^{-1}  \partial^{\nu}h_{\mu \nu} \omega(-\Box_{\Lambda}) \partial_{\rho}h^{\rho \mu}$
% \label{GF2}
 ($\omega( - \Box_{\Lambda})$ is a weight functional \cite{Stelle, Shapirobook}.)
The d'Alembertian operator in $\mathcal{L}_{\rm lin}$ and the gauge fixing term must be conceived on %relative 
%to 
the flat spacetime. 
Inverting the operator $\mathcal{O}$ \cite{HigherDG}, we find the %following 
two-point function in the harmonic gauge ($\partial^{\mu} h_{\mu \nu} = 0$),
\be
\mathcal{O}^{-1} = \frac{\xi (2P^{(1)} + \bar{P}^{(0)} ) }{2 k^2 \, \omega( k^2/\Lambda^2)} 
%&& \hspace{-0.4cm} 
+ 
\frac{P^{(2)}}{k^2   P(k^2/\Lambda^2) }
\label{propagator} 
%&& \hspace{-0.5cm} 
- \frac{P^{(0)}}{k^2  Q(k^2/\Lambda^2) \left( D-2 \right)} \,  .
\ee
%
 %\\
%&& \hspace{-0.4cm} 
%+ \frac{\zeta}{2 k^2 \, \omega_1( k^2/\Lambda^2) (2P^{(1)} + \bar{P}^{(0)}. }
%\nonumber \\
We omitted the tensorial 
indexes for the propagator $\mathcal{O}^{-1}$ and the projectors $\{ P^{(0)},P^{(2)},P^{(1)},\bar{P}^{(0)}\}$ 
%The projectors are 
%defined by  %We have also introduced the following projectors 
\cite{HigherDG, VN} %\label{proje2}
are given in the footnote\footnote{Projectors:
%%%%%
\be
 && \hspace{-1.0cm}
 P^{(2)}_{\mu \nu, \rho \sigma}(k) = \frac{1}{2}( \theta_{\mu \rho} \theta_{\nu \sigma} +
 \theta_{\mu \sigma} \theta_{\nu \rho} ) -  \frac{1}{D-1} \theta_{\mu \nu} \theta_{\rho \sigma}
 \, ,
 \,\,\,\,  \,
 %\nonumber \\
% && % \hspace{-0.2cm}
P^{(1)}_{\mu \nu, \rho \sigma}(k) = \frac{1}{2} \left( \theta_{\mu \rho} \omega_{\nu \sigma} +
 \theta_{\mu \sigma} \omega_{\nu \rho}  +
 \theta_{\nu \rho} \omega_{\mu \sigma}  +
  \theta_{\nu \sigma} \omega_{\mu \rho}  \right)  \,  , \nonumber   \\
   &&
  \hspace{-1.0cm}
P^{(0)} _{\mu\nu, \rho\sigma} (k) =  \frac{1}{D-1}\theta_{\mu \nu} \theta_{\rho \sigma}  \, ,  \,\,\,\, \,
\bar{P}^{(0)} _{\mu\nu, \rho\sigma} (k) =  \omega_{\mu \nu} \omega_{\rho \sigma} \,  , % \nonumber  \\
\,\,\,\, 
%&& %\hspace{-1.4cm}
\theta_{\mu \nu} = \eta_{\mu \nu} - \frac{k_{\mu } k_{\nu }}{k^2} \, , %\nonumber \\ 
\,\,\,\, 
%&& \hspace{-1.4cm}
\omega_{\mu \nu } = \frac{k_{\mu} k_{\nu}}{k^2} \, .
  \label{proje3}
\ee
}.
We also have replaced $-\Box \rightarrow k^2$ in the linearized action.
%where
% $\theta_{\mu \nu} = \eta_{\mu \nu} - k_{\mu } k_{\nu }/k^2$ and $\omega_{\mu \nu } = k_{\mu} k_{\nu}/k^2$.

In our construction the polynomials  %of the d'Alembertian operator 
$P(z)$ and $Q(z)$ 
can only show up complex conjugate poles and are chosen to satisfy the condition %not show real poles and 
$P(0)=Q(0)=1$. 
%This property is crucial to avoid real bound states at tree level and 
%at any order in the perturbative loop expansion {\color{red}  if the theory is finite as we will shortly show. }
%In fact, the equation $P(z)=0$ has c
The complex conjugate %pair ghost 
solutions of $P(z)=0$ and $Q(z)=0$ are ghostlike, but they %, but such pairs of 
%are complex ghosts bound states 
do not contribute to
the absorptive part of physical scattering amplitudes and may consistently be excluded from the asymptotic observable states of the theory making 
use of the Lee-Wick prescription for the construction of a unitary S-matrix over the 
physical subspace \cite{LW, Yamamoto1}. The theory is also classically stable when Lee and Wick %proposed to remove the instability by imposing 
appropriate boundary conditions are imposed \cite{Cutk,Barnaby}.

%\paragraph*{Four-dimensional theory ---} 
\subsection{Four-dimensional theory}
In $D=4$, assuming $P(z)=Q(z)$ and introducing a potential consisting only of two killer operators quartic in the curvature, the theory simplifies to 
\be
 %&& 
 %\hspace{0.1cm}
 %&& \hspace{-1cm}
%\boxed{
  \mathcal{L} 
%\bar{\lambda} - \frac{2}{\kappa_4^{2}}  R 
%\nonumber \\
%&& \hspace{-1.3cm}
%-  2 \, G_{\mu \nu} \,  \frac{ {P(-\Box_{\Lambda})} -1}{ {\kappa}_4^2 \Box}   R^{\mu \nu}  
%-  \frac{2 s_1}{ {\kappa}_4^2} \, R^2 \, \Box^{ \gamma -2} R^2 - \frac{2 s_2}{ {\kappa}_4^2}
%R_{\mu \nu} R^{\mu\nu} \, \Box^{ \gamma -2} R_{\rho \sigma} R^{\rho \sigma}
%\nonumber \\
%&& 
%\hspace{-0.65cm} 
= - 2 \kappa_4^{-2} \left[ 
- 2 \Lambda_{\rm cc} %\frac{\kappa_4^2 \bar{\lambda}}{2} 
+ R 
%\nonumber \\
%&& \hspace{-1.3cm}
-  G_{\mu \nu}   \sum_{i=0}^{ \gamma} \frac{a_i}{\Lambda^2} (- \Box_{\Lambda})^{i}
   R^{\mu \nu} 
\label{Action2b4d}
  %\nonumber\\&&
+  s_1  R^2 \, \Box^{ \gamma -2} R^2 + s_2 R_{\mu \nu} R^{\mu\nu} \, \Box^{ \gamma -2} R_{\rho \sigma} R^{\rho \sigma}  \right] \, .
\ee
In the search for a finite theory of quantum gravity, the most economic one  %theory finite at quantum level 
is obtained for $\gamma=3$ and $a_i = 0$ for $i=0,1,2$ while $a_3=1$ in (\ref{PolyG}), %adding a local potential $O(\mathcal{R}^3)$ to the action (\ref{kin4}), 
namely 
%{\color{red}If $S = \int d^4 x \, 2 \mathcal{L}/\kappa^2_4$, the Lagrangian reads
%
\be
%\boxed{
S_{\rm F} =   \int d^4 x \sqrt{-g} \, 2 \kappa_4^{-2} \left[ 
 -R + 2 \Lambda_{\rm cc} 
 - s_0 \, G_{\mu \nu}  \Box^3 %_{\Lambda}^{3} 
 R^{\mu \nu} - s_1 R^2 \Box R^2 
 - s_2
  R_{\mu \nu}^2 %R^{\mu \nu} 
  \Box R_{\rho \sigma}^2 \right]  \,    ,
  %R^{\rho \sigma}    .%\nonumber 
  \label{4Daction}
\ee
%and $s_0 \equiv 1/\Lambda^8$,
where $s_0 = 1/\Lambda^8$. 
If we are happy with super-renormalizability we can study the following minimal action,
\be 
 %\hspace{-0.6cm}
S_{\rm SR}  = \int \! d^4 x \sqrt{|g|} \, 2 \kappa_4^{-2} \Big[- R + 2 \Lambda_{\rm cc}
 - s_0 G_{\mu \nu} \Box
 R^{\mu \nu}  - \sum_i ({\bf Riem}^3)_i
 %- s_1 R^3 - s_2 R_{\mu\nu} R^{\mu \rho} R^{\nu}_{\rho} 
 \Big]  \,  ,
 \label{4Daction2}
 \ee 
where now $s_0 = 1/\Lambda^4$, and the sum is over all possible invariants cubic in the Riemann tensor
($6$ independent operators \cite{CurvatureInv3} \footnote{ 
At the cubic order in the Riemann tensor the basis of curvature invariants 
consists on eight members \cite{CurvatureInv3}, namely 
\be
%\label{eq: R3}
&&\hspace{-1cm}
%I^3_1 \equiv 
R^3 
,\quad
%I^3_2 \equiv 
R R_{\mu\nu} R^{\mu\nu} 
,\quad
%I^3_3 \equiv 
R_{\nu\alpha} R^\nu_\mu R^{\alpha\mu} 
, \\
&& \hspace{-1cm}
%I^3_4 \equiv 
R_{\nu\alpha} R_{\mu\beta} R^{\nu\mu\alpha\beta} 
,\quad
%I^3_5 \equiv 
R R_{\mu\nu\alpha\beta} R^{\mu\nu\alpha\beta} 
,\quad
%I^3_6 \equiv 
R_{\nu\alpha} R_{\beta\gamma\e}^\nu R^{\beta\gamma\e\alpha} 
, \quad %\nonumber\\&&
%I^3_7 \equiv 
R_{\mu\nu\alpha\beta} R^{\mu\nu}_{\gamma\e} R^{\alpha\beta\gamma\e}
,\quad
%I^3_8 \equiv 
R_{\mu\nu\alpha\beta} R^\mu_\gamma{}^\alpha_\e R^{\nu\gamma\beta\e} .
\label{Riem2}
\ee
but only three out of the five Riemann terms in (\ref{Riem2}) are independent in $D=4$. 
}.)
More details about the finiteness will be given later in section (\ref{QDIV}). 
%As anticipated in the introduction if the theory is defined by only one polynomial then the tensorial
%structure involve 

 %\paragraph{THE GRAVITON PROPAGATOR.} \label%The graviton propagator.}
% \section%
 %\paragraph*
 %{Propagator in momentum space}%\label{gravitonpropagator}
 
\section{Complex conjugate poles and unitarity} %Feynman \& Wheeler propagators}
We hereby study the propagator for the two minimal four dimensional theories proposed
in (\ref{4Daction}) and (\ref{4Daction2}). 
Since $P(z) = Q(z)$, the denominator of the propagator consists on the product of the monomial $k^2$ times the polynomial $P(k^2/\Lambda^2)$.
Therefore, we have the usual graviton massless pole with the same tensorial structure already found in the 
Einstein-Hilbert action, plus other complex conjugate poles resulting from the particular choice for 
the polynomial $P(z)$. 

For the theory (\ref{4Daction2}) the polynomial is $P(-\Box_{\Lambda}) = 1 +(- \Box_{\Lambda})^2 = 1+ k^4/\Lambda^4$ and the propagator in (\ref{propagator}), leaving out the gauge dependent terms, decomposes in
\be
\hspace{-0.5cm}
%&& 
\mathcal{O}^{-1}= \frac{ 1 } {k^2   P(\frac{k^2}{\Lambda^2})}
\underbrace{\left( P^{(2)} 
- \frac{P^{(0)}}{D-2 }  \right)}_{\rm TS} 
= \frac{|\eta|^4 \, {\rm TS} }{k^2(k^2 - \eta^2) (k^2 - \eta^{* 2} )}
= \left( \frac{1}{ k^2+ i \epsilon } + \frac{c^2}{k^2 - \eta^2} +  \frac{c^{* 2}}{k^2 - \eta^{* 2} } \right) {\rm TS}% \nonumber \\
%&& \hspace{0.8cm}
 \,  ,
\label{Prop2}
\ee
with 
\be
&&1+c^2 + c^{*2} = 0 \,\, , %\nonumber \\ % 
\,\,\,\, 
%&& 
c^2 \eta^2 + c^{*2 } \eta^{* 2} = 0 \,\, ,  %\nonumber \\%
 \,\,\, \, 
%&&
\eta^2 =  - i \Lambda^2 \,\, , \,\,\,\, \eta^{*2} =   i \Lambda^2 \, .
\ee
For the theory (\ref{4Daction}) the polynomial is $P(-\Box_{\Lambda}) = 1 +(- \Box_{\Lambda})^4 = 1+ k^8/\Lambda^8$ and the propagator decomposes in
\be
&& \hspace{0cm}
\mathcal{O}^{-1}= \frac{ 1}{k^2  P(k^2/\Lambda^2)}%\Lambda^8 } {k^2 \left( k^8+ \Lambda^8 \right)}
{\rm TS} 
= \frac{|\eta_1|^4 |\eta_2|^4  }{k^2 (k^2 - \eta_1^2) (k^2 - \eta_1^{* 2} ) (k^2 - \eta_2^2) (k^2 - \eta_2^{* 2} )} {\rm TS} \,  \nonumber \\
&& 
\hspace{0.8cm} 
= \left[ \frac{1}{ k^2+ i \epsilon } + \frac{c_1^2}{ k^2 - \eta_1^2 } 
+ \frac{c_1^{* 2}} {k^2 - \eta_1^{* 2} }
+ \frac{c_2^2}{ k^2 - \eta_2^2 } 
+ \frac{c_2^{* 2}} {k^2 - \eta_2^{* 2} } \right] {\rm TS} \, , 
\label{Prop1}
\ee
with complex masses square 
%1+c^2 + c^{*2} = 0 \,\, , \,\,\,\, c^2 \eta^2 + c^{*2 } \eta^{* 2} = 0 \,\, , \,\,\, \, 
\be
\eta_1^2 = e^{i \frac{\pi}{4}}  \Lambda^2 \,\, \mbox{and} \,\, \eta_2^2 = - e^{i \frac{\pi}{4}}  \Lambda^2.
\label{masse1}
\ee 
For the sake of simplicity, we considered $p(z) = 1+ z^2$ and $p(z) = 1+ z^4$, and we found the above particular values
for the masses %$\eta$, 
$\eta_1$ and $\eta_2$ (\ref{masse1}). However, following the reference %H. Yamamoto 
\cite{Yamamoto1} we can show that the group velocity $v_g$ 
for the particles with complex masses is smaller or equal then the light velocity iff the following condition are satisfied,
\be
\hspace{-0.5cm}
{\rm Re} (\eta^2) \geqslant 0 \, , \,\,\,\, {\rm Re} (\eta_1^2) \geqslant 0 \, , \,\,\,\, {\rm Re} (\eta_2^2) \geqslant 0 \, ,
\quad  
\label{groupV}
v_g = \frac{|\vec{p}|}{\sqrt{2}} \frac{\sqrt{\sqrt{(\vec{p}^2 +{\rm Re}(\eta^2))^2+({\rm Im}(\eta^2))^2 }
+ \vec{p}^2 + {\rm Re}(\eta^2))}}{\sqrt{(\vec{p}^2+{\rm Re}(\eta^2))^2+({\rm Im}(\eta^2))^2}} .
\ee
These inequalities are not met by the theory (\ref{4Daction}), but are perfectly satisfied by 
the minimal super- renormalizable theory (\ref{4Daction2}). %for the choice of the polynomial $P(z)$
%, but the expressions for the propagators above in (\ref{Prop2}) and (\ref{Prop1}) are correct. 
If we want (\ref{groupV}) to be fulfilled for complex conjugate pairs with a strictly positive real part of the mass square, %group velocity 
we have just to replace the polynomial in (\ref{Prop2}) with the following one,
%reconstruct the right polynomial compatible with finite group velocity. % in oder to fulfill (\ref{groupV}). 
%For the propagator in (\ref{Prop2}) the polynomial is:
\be
 P(k^2/\Lambda^2) = \frac{k^4 - k^2 {\rm Re}(\eta^2) + |\eta|^4}{|\eta|^4}= 
 \frac{\Box^2 + \Box \, {\rm Re}(\eta^2) + |\eta|^4}{|\eta|^4} \, ,
 \label{PY}
 \ee
and the theory reads,
\be
S_{\rm SR}^\prime \! = \!\!  \int \! d^4 x \sqrt{|g|} 2 \kappa_4^{-2} \Big[ - R + \bar{\lambda}
 - G_{\mu \nu} \frac{\Box + {\rm Re}(\eta^2) }{|\eta|^4}  R^{\mu \nu} 
- \sum_i ({\bf Riem}^3)_i
% + s_1 R^3 + s_2 R_{\mu\nu} R^{\mu \rho} R^{\nu}_{\rho} 
\Big] .
 \label{SY}
\ee
We can also make (\ref{4Daction}) compatible with (\ref{groupV}) by replacing the polynomial with the following one,
\be
P(k^2/\Lambda^2) = \frac{k^8 + k^4 m_1^4 + k^4 m_2^4  +m_1^4 m_2^4}{\Lambda^8}  
= \frac{(-\Box)^4 + (-\Box)^2 m_1^4 + (- \Box)^2 m_2^4  +m_1^4 m_2^4}{\Lambda^8} 
\,  ,
\ee
where $m_1^4 m_2^4 = \Lambda^8$ and $m_1,m_2 \in \mathbb{R}$.
The action now reads,
\be
 {S_{\rm F}^{\prime}} =   \int d^4 x \sqrt{-g} \, 2 \kappa_4^{-2} \left[ 
 -R +\bar{\lambda} 
 - s_0 \, G_{\mu \nu}  (\Box^3 + \Box \, m_1^4 + \Box \, m_2^4)
 R^{\mu \nu} - s_1 R^2 \Box R^2 - s_2
  R_{\mu \nu}^2 %R^{\mu \nu} 
  \Box R_{\rho \sigma}^2 \right]   ,
  %R^{\rho \sigma}    .%\nonumber 
  \label{4DactionY2}
\ee
where again $s_0 = 1/\Lambda^8$ and the poles are now located in:
%\be
$\left\{-i m_1^2,\, i  m_1^2, \, -i m_2^2  , \,  i m_2^2\right\}$. %\quad \Longrightarrow \quad v_g = 0 \, . 
%\ee
All the complex poles in (\ref{4DactionY2}) have group velocity 
zero (and real part of the mass square zero) like for the theory (\ref{4Daction2}). 
We can get positive group velocity taking the following polynomial,
\be
&& \hspace{-1.55cm} 
P(k^2/\Lambda^2) = \Lambda^{-8} \, \left( 
k^8-k^6 m_1^2-k^6 m_2^2+k^4 m_1^4+k^4 m_2^4+k^4 m_1^2 m_2^2-k^2 m_1^2 m_2^4-k^2 m_1^4
   m_2^2+m_1^4 m_2^4 \right) %}{\Lambda^8}  
   \\
   && 
   \hspace{0.05cm} 
   =\Lambda^{-8} \, \left( 
\Box^4 + \Box^3 m_1^2 + \Box^3 m_2^2+ \Box^2 m_1^4+ \Box^2 m_2^4+ \Box^2 m_1^2 m_2^2
+\Box m_1^2 m_2^4 +\Box m_1^4
   m_2^2 \right)  + 1\, , \nonumber 
\ee
and the complex conjugate poles are now located in:
\be
\left\{  \frac{1}{2} \left(1-\sqrt{3} i\right) m_1^2 , \,  \frac{1}{2} \left(\sqrt{3} i+1\right)
   m_1^2 , \,  \frac{1}{2} \left(1-\sqrt{3} i\right) m_2^2 , \,  \frac{1}{2}
   \left(\sqrt{3} i+1\right) m_2^2   \right\} \, . %\quad \Longrightarrow \quad v_g = \frac{1}{2} \, . 
   \ee

 % =\frac{
%(- \Box)^4- (-\Box)^3 m_1^2- (-\Box)^3 m_2^2+ (-\Box)^2 m_1^4+ (-\Box)^2 m_2^4+ (- \Box)^2 m_1^2 m_2^2
%- ( - \Box) m_1^2 m_2^4 - (- \Box) m_1^4
 %  m_2^2}{\Lambda^8}  + 1 

Now we would illustrate more in detail the unitarity of the proposed actions.  
The theories under consideration are marked by pairs of complex conjugate poles.
In (\ref{4Daction2}) we have one pair of complex conjugate poles, while in 
(\ref{4Daction}) we  have two complex conjugate poles, etc. 
We discarded extra real particles from the spectrum of the classical theory, but we allow for 
conjugate pairs of unstable and unphysical particles: ``anti-gravitons". 
It is well known that, at least for a single pair of complex conjugate poles, 
a unitary S-matrix defined between physical asymptotic states exists \cite{LW, Cutk}.
The unphysical particles do not contribute to the absorptive part of the 
propagators (\ref{Prop2}) or (\ref{Prop1}) because they occur as complex conjugate pairs.
We can easily check that the complex conjugate poles do not go on shell by taking the imaginary part of any one of the propagators above, namely 
\be
{\rm Im} (\mathcal{O}^{-1}(k) ) =  - \frac{\epsilon}{k^4 + \epsilon^2} \,\, \rightarrow \,\, -  \pi \,  \delta^4(k^2) \, . 
\label{NoGoOnShell}
\ee
%Therefore the complex conjugate poles do not go on shell.
%
Since the incoming particles have real energy and momentum they can not produce on-shell intermediate states with complex mass. Therefore 
the complex poles do not destroy unitarity and 
%, and therefore we can not produce on shell intermediate states with complex mass. 
their contribution to the scattering amplitudes is real. Indeed we can easily verify that the tree level exchange satisfies the optical theorem as a consequence of the energy momentum conservation that generally  follows from the definition of the S-matrix. 
From unitarity follows that the imaginary part of the forward scattering amplitude, $\mathcal{M}$, must be a positive quantity (optical theorem.) For example the inequality $2{\rm Im}[\mathcal{M}(2,2)]>0$ is satisfied at tree level 
as a mere consequence of (\ref{NoGoOnShell}) 
\cite{LW,Jansen,Yamamoto1,Veltman}. Only the massless gravitons contribute to the imaginary part of the amplitude, while
the anti-gravitons give contribution to the real part making it more convergent in the ultraviolet regime. 
Since (\ref{NoGoOnShell}) the tree-level unitarity for every propagators obtained in this section reads as follow,
\be 
 2 \, {\rm Im} \left\{  T(k)^{\mu\nu} \mathcal{O}^{-1}_{\mu\nu, \rho \sigma} T(k)^{\rho \sigma} \right\} = 2  \pi \, {\rm Res} \left\{  T(k)^{\mu\nu} \mathcal{O}^{-1}_{\mu\nu, \rho \sigma} T(k)^{\rho \sigma} \right\} \big|_{k^2 = 0}> 0.
 \ee
 where $T_{\mu\nu}(k)$ is the conserved energy-momentum tensor. 

At quantum level the theory can be super-renormalizable or finite
(see section five for more details.) 
% if a proper curvature potential is introduced.
For the sake of simplicity and strictness, let us start considering the case of a finite theory. For this class of theories the beta functions are zero, we do not have to introduce counterterms,
the propagator does not change (for what about divergences), 
and so the Lee-Wick unitarity is safe. 
However, we of course have finite contributions to the quantum effective action at any order in the loop expansion.
%The effective action is not unique and nonlocal analytic contributions can often be removed by a field redefinition. %\footnote{We can study the one loop correction to the propagator coming from the ...}.
Nevertheless, %when operators quadratic in curvature can non be converted in vertexes, 
at perturbative level we typically have 
a slight displacement in the position of the complex conjugate poles or in the worst case a larger 
number of them up to infinity depending on the peculiar finite quantum nonlocal 
contributions to the effective action. Therefore, the unitarity Lee-Wick structure of the classical theory is 
likely preserved at quantum level. 

For super-renormalizable theories we have logarithmic divergences and the running of the coupling 
constants comes along with the following nonlocal operators in $D=4$,
\be
\alpha_1 \, 
\mathcal{R} \log\left(  \frac{-\Box}{\mu^2} \right) \mathcal{R}  .
\label{Q4}
\ee
In $D=5$ the theory is finite and we expect the following contribution, 
 \be
\alpha_2  \, 
\mathcal{R} \sqrt{   -\Box  }\, \mathcal{R}  .
\label{Q5}
\ee
Therefore, we end up with a quantum theory having the same structure of the initial classical theory, but  
a shift in the position of the complex conjugate poles, or, in the worst case, more complex conjugate poles that we can easily handle applying again the Lee-Wick prescription at the tree-level quantum action. 
The quantum corrections to the spin two and/or spin zero inverse propagators (\ref{Prop2}) implied by (\ref{Q4}) and (\ref{Q5}) respectively read
\be
k^2 \left( 1+ \frac{k^4}{\Lambda^4} + \alpha_1 \kappa_4^2 \, k^2 \, \log \frac{ k^2}{\mu^2}   \right) \,\,\,\, {\rm or}
\,\,\,\, \,\, 
k^2 \left( 1+ \frac{k^4}{\Lambda^4} + \alpha_2 \kappa_4^2 \, k^2 \, \sqrt{ k^2}  \right).
\ee
It is straightforward to cheek that the number of complex conjugate poles do not change, even though 
they are slightly moved out from the original classical position. Once again, unitarity 
is not affected by the quantum corrections. 

For the special super-renormalizable theory (\ref{4Daction2}) with propagator (\ref{Prop2}) there are only 
two complex conjugate poles, therefore, we can apply all the results derived in the paper \cite{Cutk}. 
In particular in \cite{Cutk} it is given a proof of perturbative unitarity compatible with Lorentz invariance
exploring a large class of Feynman diagrams, while the {\em acausal} effects are fantastically small to be 
detected. In \cite{Wise3} it is given an explicit proof of the one-loop unitarity based on the CLOP \cite{Cutk}  prescription to integrate in the complex energy plane. In \cite{Liu} it is given a different and likely unambiguous Feynman 
$i\epsilon$ prescription. 
However, the formalism developed in \cite{BV0, BV, V2}
for any higher derivative theory produces a unique quantum effective action and the above ambiguities 
are avoided. The quantization of a general gravitational theory with an arbitrary number of complex conjugate poles has to be understood as a mere application of the 
procedure explained in section 2.1 of the report \cite{BV}.

%Throughout this section we made tacit use of 
%the background field method (BFM), Barvinsky-Vilkovisky techniques (BV), and Batalin-Vilkovisky formalism. 
%Actually,  BFM and BV are crucial tools in making the %, therefore our claims 
%about 
%unitarity claim stable at any perturbative order. The same arguments may be proved 
%making use 
%using Feynman diagrams in flat spacetime, but with a much greater effort. 
In short, the Lee-Wick unitarity reads as follows:
the S-matrix is unitary in the physical subspace of real states (only gravitons in this section), 
while complex mass particles appear only as virtual states. 
Assuming the Lee-Wick \cite{LW} or \cite{Yamamoto1} definitions, the S-matrix vanish for all non real initial and final states, while the unphysical complex states can appear only as virtual states. 
Therefore, the S-matrix is unitary as a mapping in the subspace of real physical states. 
The complex poles occur in a proper combination %, %and do not raise any divergence in the 
%Feynman propagator for $t\rightarrow \pm \infty$ respecting macrocausality as defined in \cite{Yamamoto1}.
%At the same time they 
to cancel out the divergences that arise from the physical states. 
Once again, complex particles are consistently excluded from the asymptotic states 
preserving the usual unitarity notion in the subspace of real states. 
%, but all the complex pairs (anti-graviton conjugate couples), 
%one for each mode $k$, occur in loops making them more convergent. 
%expanded by the physical state vectors.
%
%
%
The Hamiltonian approach remains well defined in the indefinite metric Hilbert space \cite{LW, Jansen}.
%\footnote{
%For the simpler case of two complex conjugate poles and a massless particle the Hamiltonian read,
%\be
%H = 
%\ee
%where ..}. 
%
%
The theory is unitary and Lorentz invariant, but microcausality is violated %as in any nonlocal theory
\cite{LW, Cutk, Tonder}. However, macrocausality is preserved because the Feynman propagator is convergent in the limit $|x_0| \rightarrow +\infty$, namely the propagator does not diverge for infinite separation time, as easy to see making the explicit integration in the energy complex plane \cite{Yamamoto1}: 
%
%\be
$\big| \big| \langle 0 |  T(h_{\mu\nu}(x) h_{\rho \sigma}(y) | 0 \rangle \big| \big| < \infty$ % \quad {\rm for} 
%\quad  
for $|x_0| \rightarrow +\infty$. 
%\ee

\section{Propagator in coordinates space and gravitational potential}
{\em Propagator in coordinates space ---}
We can easily obtain the propagator in coordinate space by the Fourier transform 
of (\ref{propagator}).
%In this section we give the propagator in coordinate space and 
%we assume that the theory is renormalized at some scale $\mu_0$; therefore, 
%If we want that 
Let us consider the more convergent case $P(z)=1+z^4$ and omit the tensorial structure in 
(\ref{propagator}), then the propagator in coordinate space reads
\be
%&& \hspace{-0.5cm} 
G(x-y) = %\\%&&  \hspace{-0.5cm} 
 \frac{1}{4 \pi^2 (x-y)^2}- \frac{ G_{0,8}^{5,0}\left(\frac{(x-y)^8}{16777216} \Big|
\begin{array}{c}
 -\frac{1}{4},0,\frac{1}{4},\frac{1}{2},\frac{3}{4},0,\frac{1}{4},\frac{1}{2} \\
\end{array}
\right)}{(16 \pi)^2} \, , \,\,\,
 G(0) = \frac{\pi }{64 \sqrt{2} \Gamma \left(\frac{1}{4}\right) \Gamma
   \left(\frac{3}{4}\right)^2 \Gamma \left(\frac{5}{4}\right)} .
%\nonumber 
\ee
In the coincidence limit the two point function does not diverge approaching the constant %, %namely
%\be
$G(0)$. % = \frac{\pi }{64 \sqrt{2} \Gamma \left(\frac{1}{4}\right) \Gamma
 %  \left(\frac{3}{4}\right)^2 \Gamma \left(\frac{5}{4}\right)}$.
 %  \ee
%\be 
%&& 
%G(x ) = \frac{\pi }{64 \sqrt{2} \Gamma \left(\frac{1}{4}\right) \Gamma
%   \left(\frac{3}{4}\right)^2 \Gamma \left(\frac{5}{4}\right)} % \nonumber \\&& 
%   -\frac{\pi  \, x^2}{4096
 %  \Gamma \left(\frac{3}{4}\right)^2 \Gamma \left(\frac{5}{4}\right)^2} + O((x)^2) \, . 
%\ee
The high energy behaviour of the two point function here derived has a universal character. Indeed, whatever the polynomial (or non-polynomial \cite{modesto}) form factor is, the short distance limit always attains a constant value. 

{\em Gravitational Potential ---}
%In this section we calculate the gravitational potential for different local form factors.\\
%
To address the problem of classical singularities
%mentioned at the beginning of the paper,  
%the first quantity to calculate in a general modified theory
%of gravity is 
we can begin by calculating the Newtonian gravitational potential. 
Given any propagator, the graviton solution of the linear equations of motion %resulting 
%from the Lagrangian (\ref{LGM}) with $g= \kappa/2$, 
is: 
\be
%&& 
\hspace{-0.6cm} h_{\mu \nu}(x) = \frac{\kappa_D}{2} \int d^D x^{\prime} 
\mathcal{O}^{-1}_{\mu \nu, \rho \sigma}(x-x^{\prime}) T^{\rho \sigma} (x^{\prime}) \label{hmunuD} % \\
%&& \hspace{-0.6cm}
 ={\frac{\kappa_D}{2} }\int \! d^D x^\prime \! \! \int  \! \frac{d^D k}{(2 \pi)^D} 
 \frac{e^{ i k (x - x^\prime)} }{ k^2 P(k^2/\Lambda^2)   }%\nonumber \\
%&& \hspace{0.9cm} 
\left(T_{\mu \nu} - \frac{\eta_{\mu \nu} }{D-2} T^\mu_\mu \right) %e^{- H(k^2/\Lambda^2)}}
 .
%\left(T_{\mu \nu} - \frac{\eta_{\mu \nu} T }{D-2} \right).
\ee
%where the graviton field is now dimensionless. 
For a static source with energy tensor 
%\be
$T^{\mu}_{\nu} = {\rm diag}(M \, \delta^{D-1}(\vec{x}), 0, \dots, 0)$, 
%\label{staticSource}
%\ee 
the spherically symmetric solution %of (\ref{hmunuD}) satisfying spherical symmetry
reads, 
\be
%&&
\hspace{-0.7cm}
 h_{\mu \nu}(r) = - \frac{\kappa_D M}{2} E_{\mu \nu} \int \frac{d^{D-1} k}{(2 \pi)^{D-1}} \, 
\frac{e^{- i \vec{k} \cdot \vec{x}}  }{\vec{k}^2 \,   
P(\vec{k}^2/\Lambda^2)
}   \label{hd}
%\\&& \hspace{-0.5cm}
=  - \frac{\kappa_D M}{2} \, \frac{ \pi^{ \frac{D-3}{2} }}{(2 \pi)^{D-2} } \, 
 \frac{E_{\mu \nu}}{r^{D-3}} %\times \nonumber \\
%&&  \hspace{-0.6cm} \times 
\int d p \, \frac{p^{D-4} \,_0\tilde{F}_1\left( \frac{D-1}{2} ; - \frac{p^2}{4} \right)}{P(p^2/r^2\Lambda^2)}
  , 
 \ee
where 
$\,_0\tilde{F}_1(a;z) = \,_0{F}_1(a;z)/\Gamma(a)$ is the regularized hypergeometric confluent function. 
In (\ref{hd}), we also have introduced 
the variable $p = |\vec{k}| r$ and the matrix
$E_{\mu \nu} = (D-2)^{-1}{\rm diag}\left(D-3,1 , \dots, 1 \right).$
Using the graviton solution above in (\ref{hmunuD}, \ref{hd}) we can reconstruct all the components of the metric tensor and then we can get the spacetime line element for a spherically symmetric source. 
The gravitational potential is related to the $h_{00}$ component 
of the graviton field by $\Phi = \kappa_D h_{00}/2$. Then, %using for a static source %with spherically symmetric 
using (\ref{hd}) %for a static %with energy tensor 
%$T^{\mu}_{\nu} = {\rm diag}(M \, \delta(\vec{x}), 0, \dots, 0)$, 
we get 
\be
&& \hspace{-1.0cm}
%\hspace{0.0cm}
\Phi(r) %\frac{\kappa^2}{4} \frac{D-3}{D-2} 
%\int d^D x^\prime \int \frac{d^D k}{(2 \pi)^D} e^{ i k (x - x^\prime)} \nonumber \\
%&& \hspace{0.8cm} \frac{e^{- H(k^2/\Lambda^2)}}{ k^2} M \int \frac{d^{D-1} \vec{q}}{(2 \pi)^{D-1}} \, 
%e^{i \vec{q} \cdot \vec{x}^{\prime}} \nonumber \\
%&& \hspace{-0.5cm} 
= - \frac{\kappa_D^2 M }{4} \frac{D-3}{D-2} \int \!\! \frac{d^{D-1} k}{(2 \pi)^{D-1}} 
\frac{ e^{ - i \vec{k} \cdot \vec{x}}}{ \vec{k}^2  \, P \left( \vec{k}^2/\Lambda^2 \right)}
%\nonumber \\
%&& \hspace{0.0cm}
= - \frac{G_N M}{r^{D-3}} \, 2 \frac{D-3}{D-2} \, \frac{ 2^{4-D} }{\pi^{\frac{D-3}{2}}} \,
%\, 
% \frac{2^{4-D}}{\Gamma \left(\frac{D-3}{2} \right)} \!
\int \! dp %\, p^{D-4} 
\, \frac{ \,_0\tilde{F}_1\left( \frac{D-1}{2} ; - \frac{p^2}{4}  \right) }{p^{4 -D} \, P\left( p^2/r^2 \Lambda^2 \right)  } 
%\mathbb{F}_D(r)
 \, .
%\nonumber  \\
%&& %\hspace{-1cm}
%\ee
%where the function $\mathbb{F}_D(r)$ is defined by 
%\be
% \mathbb{F}_D(r) \equiv \frac{2^{4-D}}{\Gamma \left(\frac{D-3}{2} \right)} \!
%\int \! dp %\, p^{D-4} 
%\, \frac{ \,_0\tilde{F}_1\left( \frac{D-1}{2} ; - \frac{p^2}{4}  \right) }{p^{4 -D} \, P\left( \frac{p^2}{r^2 \Lambda^2} \right)  } %\nonumber \\
%&& \hspace{3cm}
\label{pote0} %\label{FD}
\ee
%\hspace{-0.35cm}
For example, in $D=4$, %we can calculate numerically 
(\ref{pote0}) simplifies to  
%Introducing again the variable 
%$p = |\vec{k}| r$ in spherical coordinates, we get 
\be
\Phi(r) = - \frac{G_N M}{r}\frac{2}{\pi} \int_0^{+\infty} d p \, \frac{J_0( p )}{{P \left( p^2/r^2 \Lambda^2 \right) }} \, %\frac{\sin(p)}{p} 
 \, , \,\,\,\, J_0(p) = {\rm sinc}(p) \equiv \frac{\sin( p )}{p} \, . 
\label{pot4D}
\ee
%and (\ref{pot4D}) can be integrated numerically.
%In %(\ref{pot4D}) 
%this latter, 
%where $J_0(p) = {\rm sinc}(p) \equiv \sin(p)/p$ is the Bessel function. 
%
%
%%%%%%%
In the table below we explicitly give the potential for three different selected polynomial 
$P(z)$. % (Stelle's theory)
%\subsection{Towards Black Hole Solutions}
	%WE CAN SUMMARIZE HERE THE RESULT FOUND IN THE OTHER PAPER 

%\subsection{Avoiding Classical and Quantum Singularities: Self-completeness}
%The properties studied above show that the gravitational force disappearance at short 
%distance. This is a characteristic feature of any asymptotically free theory. Here, asymptotic freedom 
%is achieved introducing in the classical action a finite (or infinite in the non-polynomial case) number of higher order operators regulated by an invariant mass scale $\Lambda$. 

%\hspace{-0.01cm}

%%%%%%%%%%%%%%%%%%%%%%%%%%%%%%%%%%%%%%%

\begin{center}
\begin{tabular}{r|r}
\hline
\,
${\rm Polynomial}$ \,\,\, %& $z$ \,\,\,\, 
& \, Potential   \\
\hline
\hline
\, $P(z)=1+z$ (quadratic gravity) \,   
%& \, 1.73 \,
 & $\Phi( r ) =  -\frac{m (1 - e^{- \Lambda r })}{r}$ \\
\hline
%\hline
\, $P(z)=1+z^2$ (super-renormalizable theory)\,  
%& \, 1.62 \, 
& $\Phi( r ) = - \frac{m \left(1- e^{-\frac{\Lambda r}{\sqrt{2}}} \cos \left(\frac{\Lambda r}{\sqrt{2}}\right)\right)}{r}$
%\vspace{0.1cm}
 \\
\hline
%\hline
\, $P(z)=1+z^4$ (finite theory)\,
% & \, 1.92 \, 
& 
$\Phi( r ) =  -\frac{m \left(   4- e^{- \eta_1  r} - e^{- \eta_2^{*}  r} - e^{ \eta_2  r} - 
e^{ \eta_1^{*}   r}  \right)}{4 r} $
 \\
\hline
\end{tabular}
\label{good}
\end{center}
%%%%%%%%%%%%%%%%%
For the first choice the potential is regular in $r=0$ thanks to the real ghost pole
in the propagator \cite{tiberio}, for the other choices at short distance ($\sim 1/\Lambda$) the anti-gravitons 
screen the anti-screening effect of the gravitons. 
For the case of $P(z)=1+z^4$ we can rewrite the potential in the table above in the following explicitly real form,
\be
  \Phi ( r )= - \frac{m}{r}
+
\frac{m e^{-\Lambda r \sin \left(\frac{\pi }{8}\right)} \cos \left(\Lambda r \cos \left(\frac{\pi }{8}\right)\right)}{2
   r} 
   +\frac{m e^{- \Lambda r \cos \left(\frac{\pi }{8}\right)} \cos \left(\Lambda r \sin \left(\frac{\pi
   }{8}\right)\right)}{2 r} \, .
   \ee
The reader can easily recognize the complex conjugate mass poles in the classical gravitational potential
(see the last row of the table.) 
They clearly play a crucial rule in making the potential singularity free in 
%,
%namely
%\be
%&& m_1^2 = (-1)^{3/4} \, ,  \,\, \, (m_1^2)^{ *} =- (-1)^{1/4}  \,  , \,\,\,  %\nonumber \\
 %m_2^2 = - (-1)^{3/4} \, ,  \,\,\,  (m_2^{2})^{ *} = (-1)^{1/4} \, .
% \ee
%There are conjugate double poles in 
agreement with the Lee-Wick requirement for a consistent 
theory at classical and quantum level. This result is in agreement with the interpretation 
given in a previous work \cite{tiberio}. Actually this is a generalization 
of the result in \cite{tiberio} 
to a theory with complex conjugate poles. 

In the same approximation we can reconstruct the metric for black hole or cosmological solutions \cite{ModestoMoffatNicolini, BambiMalaModesto2, BambiMalaModesto, calcagnimodesto}. Exact solutions can be found closely following the derivations in \cite{Bis3, koshe1, koshe2, koshe3}.
%There are many similarities between nonlocal theories and local higher derivative Lee-Wick theories.
%In particular many exact or approximate classical solutions obtained for nonlocal theories can be 
%easily rec

\section{Quantum Divergences}\label{QDIV}
%
%\paragraph{POWER COUNTING OF LOOP DIAGRAMS.} %Power counting of loop diagrams.}
%\paragraph*
%{\em Quantum divergences ---} %Power counting of loop diagrams}% and renormalization.} 
%\subsection{Power counting of loop diagrams}
Let us then examine the ultraviolet behaviour of the quantum theory and what kind of operators in the action
are source of divergences.
In the high energy regime
the graviton propagator in momentum space for the theory (\ref{gravityG}) 
%and the leading interaction vertex are 
schematically scales as %given by
\be
\mathcal{O}^{-1}(k) \sim \frac{1}{k^{2 \gamma +D}}.
\ee
Since the interactions have leading ultraviolet scaling $k^{2 \gamma + D}$,
we find the following upper bound to the superficial degree of divergence in a $D$-dimensional 
spacetime, 
\be
%&& 
&&%\hspace{-0.5cm} 
\omega(G) =  D - 2 \gamma  (L - 1) \, .
\label{even}
\ee
In (\ref{even})  we used the topological relation between vertexes $V$, internal lines $I$ and 
number of loops $L$: $I = V + L -1$. %For the choice we have already  introduced in the previous 
%section 
%We see that only 1-loop diagram are divergent if $\gamma > 2$ and the theory is super-renormalizable.
Thus, if $\gamma > D/2$ only 1-loop divergences survive in this theory, therefore, it is 
super-renormalizable. 
%
%\footnote{
%A ``local" super-renormalizable quantum gravity with a large 
%number of metric derivatives was for the first time introduced in \cite{shapiro}.
%}.
%\begin{widetext}
Only a finite number of constants is renormalized in the action (\ref{gravityG}), i.e. 
$\kappa_D$, $\bar{\lambda}$, $a_n$, $b_n$ together with the finite number of couplings that multiply the operators $O({\mathcal{R}}^3)$ in 
the first line of (\ref{gravityG}) up to ${\mathcal{R}}^{D/2}$. % of the action. %(\ref{action}) 
 %the quantities 
%$\beta$, $\beta_2$, $\beta_0$ and eventually the cosmological constant are renormalized, 

%%%%%%%%

%%%%%%%%
Let us now expand on the one-loop divergences. 
The main divergent integrals contributing to the one-loop effective action have the following form,
\be 
\hspace{-0.4cm}
\int \!\! \frac{d^D k}{(2 \pi)^D}  \left\{ \prod_{i=1}^{s} 
\frac{1}{(k+p_i)^{2 n}} \right\} \! P(k)_{2 s  n}  . %\nonumber \\
%&&\hspace{-5cm}
\label{integral}
\ee
$P_{2sn}(k)$ is a polynomial function of degree $2 n s$ in the momentum $k$ 
(generally it also relies on the external momenta $\bar{p}_a$), %$a=1, \dots,s$), 
$p_i = \sum_{a=1}^i \bar{p}_a$, and $n = \gamma +{\rm N} +2$ for the graviton field $h_{\mu\nu}$.
We can write, as usual,
\be
&& %\hspace{-1.5cm}
\prod_{i=1}^{s} 
\frac{1}{ (k+p_i )^{2 n}}  %= \\ && \hspace{-0.5cm}
\propto \int_0^1 \! \left( \prod_{i=1}^s x_i^{n-1} d x_i \!  \right) % \!\!
 \frac{\delta\left( 1 - {\sum_{i=1}^s x_i} \right)}{[k^{\prime 2} + \mathcal{R} ]^{n s}}  \, , \,\,\,\, 
 %\nonumber\\
% && % \hspace{-0.5cm} 
 k^{\prime} = k + \! \sum_{i=1}^s x_i p_i   \, , \,\,\,\,  %\nonumber \\
% && 
 \mathcal{R} = \sum_{i=1}^s p_i^2 x_i 
 - \left( \sum_{i=1}^s x_i p_i \right)^{\!\! 2} \!\!  \, . %\nonumber  
\ee
%where ${\rm c} = {\rm constant}$. 
%%%%%
%%%%%
In (\ref{integral}), we move outside the convergent integral in $x_i$ and we replace 
%the integration variable 
$k^{\prime}$ with $k$ %and we turn out with
%introduce the polynomial 
\be
\int \!\! \frac{d^D k}{(2 \pi)^D} 
\frac{P^{\prime}(k, p_i, x_i)_{2 s  n} }{( k^2 + \mathcal{R} )^{n s }} \, .
\label{int2}
\ee
Using Lorentz invariance and missing the argument $x_i$, we replace the polynomial 
$P^{\prime}(k, p_i, x_i)_{2 n s}$ with a polynomial of degree $n \times s$ in $k^2$, 
namely $P^{\prime \prime}(k^2, p_i)_{n s}$. %.
%In the intermediate steps we integrate (\ref{int2}) 
%in $|k|$ from zero to a cut-off $\Lambda_{\rm c}$ and then
%we take the limit $\Lambda_{\rm c} \rightarrow \infty$.  
Therefore, the integral (\ref{int2}) reduces to
\be
\int \!\!  \frac{d^D k}{(2 \pi)^D} %{\sum_{j=0}}^{\prime}  c_{u_j} 
\frac{P^{\prime \prime}(k^2, p_i)_{n s} }{( k^2 + \mathcal{R} )^{n s }} \, .
\label{int3}
\ee
We can decompose the polynomial $P^{\prime \prime}(k^2, p_i)_{n s} $ 
in a product of external and internal momenta
only to obtain the divergent contributions,
\be
&& \hspace{-0.4cm} 
P^{\prime \prime}(k^2, p_i)_{n s} = \sum_{\ell=0}^{[ D/2 ]} \alpha_{\ell}(p_i) k^{2 n s-2 \ell} = 
k^{2 n s} \alpha_0 + k^{2 n s-2} \alpha_1(p_i) + k^{2 n s - 4 } \alpha_2(p_i)   
%+ k^{2 n s - 6 } \alpha_3(p_i) 
+ \dots \, . \label{expa}
\ee
Given the polynomial 
\be
P(z) = 1+ c_{\gamma +{\mathrm N} + 1} z^{\gamma +{\mathrm N} + 1} + c_{\gamma +{\mathrm N} } z^{\gamma +{\mathrm N} } 
+c_{\gamma +{\mathrm N} -1} z^{\gamma +{\mathrm N} -1} + \dots,
\ee
we find the following logarithmic divergences, 
\be
&&\hspace{-1cm}
 \sum_{\ell=0}^{[ D/2 ]} 
\int \!\!  \frac{d^D k}{(2 \pi)^D} 
\frac{ \alpha_{\ell}(p_i) k^{2 n s-2 \ell}  }{( k^2 + \mathcal{R} )^{n s }} = \nonumber \\
&&\hspace{-1cm}
\label{int4} 
 =  \sum_{\ell=0}^{[ D/2 ]} 
  \frac{i \alpha_{\ell}(p_i) (\mathcal{R})^{\frac{D}{2}- \ell }}{(4 \pi)^{\frac{D}{2}}} \, \frac{\Gamma \left(\ell - \frac{D}{2}  \right) 
  \Gamma \left(ns -\ell + \frac{D}{2}  \right)}{\Gamma \left( \frac{D}{2}  \right) \Gamma(n s)}  %\propto 
  \,\,\,\, \Longrightarrow \,\,\,\, 
 \overbrace{ \frac{1}{\epsilon} \beta_{\lambda}\sqrt{|g|} + \frac{1}{\epsilon} \beta_R R + \dots 
  +  \frac{1}{\epsilon} \beta_{{\mathcal R}^{D/2}}\mathcal R^{D/2} }^{\rm counterterms} \, .
\ee
We schematically listed above the counterterms and explicitly introduced the ultraviolet cut-off $\epsilon$ in the dimensional regularization scheme. 

Now we specify the above general analysis to our particular class of theories (\ref{4Daction}),(\ref{4Daction2}), and (\ref{SY}). 
The three theories are super-renormalizable, but for the case of (\ref{4Daction})
we only have one loop divergences, while for (\ref{4Daction2}) and (\ref{SY}) we also have divergences 
at two loops or three loops. 

Given the particular choice of the polynomial $P(z)$ for the theory (\ref{4Daction}), 
the conterterms can only be proportional to $R^2$ and $R_{\mu\nu}^2$. 
Moreover, it is always possible to tune the front coefficients 
$s_1$ and $s_2$ for the quartic operators $\mathcal{R}^2 \Box \, \mathcal{R}^2$ and 
to make zero the beta functions.
This is due to the linearity in $s_1$ and $s_2$ of the beta functions $\beta^{R^2}$ and $\beta_{R_{\mu\nu}^2}$. 
%because their contributionto the divergences can only be linear 
For the theory (\ref{SY}) we also expect the beta functions $\beta_{R^2}$ and 
$\beta_{R_{\mu\nu}^2}$ to be zero for some special choice of the front coefficients $s_1$ and $s_2$.
However, the beta function are probably quadratic in $s_1$ and $s_2$ and only an explicit computation 
could confirm this property. At two loops we can have counterterms proportional to the Ricci scalar (Einstein-Hilbert operator) or the cosmological constant, while at three loops we only have divergences proportional to the cosmological constant.

\section{New Lee-Wick standard model} %Gauge and matter sectors}
In the previous papers \cite{LeslawModestoFinite, PivaFGT} an higher derivative and weakly nonlocal
theory beyond the standard model of particle physics has been proposed.
However, such theory is quasi-polynomial in many respects and it is straightforward to take into account 
the results in \cite{LeslawModestoFinite, PivaFGT} to propose here a local higher derivative 
and super-renormalizable action for gauge interactions and matter. 
This is a forced extension beyond the standard model if we want to preserve 
super-renormalizablity of the gravitational interactions after coupling to matter. 
Moreover, 
Lee-Wick gauge interactions turn out to be (super-)renormalizable or finite regardless of the spacetime dimension. Following the notation of section (\ref{TheTheory}), the action for gauge bosons reads as follows, 
\be \hspace{-0.2cm} 
\mathcal{L}_{\rm gauge} = -\frac{1}{4g^2}\Big[ F_{\mu\nu} P_g({\cal D}_\Lambda^2)
F^{\mu\nu} + \frac{s_g}{\Lambda^4} F^2({\cal D}_{\Lambda}^2)^{2} F^2 \Big],
\label{gauge}
\ee
where the polynomial $P_g({\cal D}_\Lambda^2)$, as a function of the square of the gauge covariant derivative ${\cal D}$, must be chosen having only complex conjugate poles and the same asymptotic behaviour as the analogue functions introduced for the pure gravity sector.
For the fermionic and scalar sectors we achieve super-renormalizability with the following action,
\be
&&  \hspace{-0.3cm}
\mathcal{L}_{\rm F} = 
 \sum_{a}^{N_f}\bar \psi_a \, i \slashed{\cal D}_{a} P_f({\cal D}_\Lambda^2)
\, \psi_a , \label{FS} \\
%+\sum_{R}\bar\psi_R \, i\slashed{\cal D}_{R} e^{H(-\slashed{\cal D}^{2}_{R,\Lambda})} 
%\psi_R 
&& \hspace{-0.3cm}
\mathcal{L}_{\rm H} =  ({\cal D}_{\mu} \Phi)^\dagger P_s({\cal D}_\Lambda^2) ({\cal D}^{\mu} \Phi) 
- \mu^2 \Phi^\dagger  P_s({\cal D}_\Lambda^2)   \Phi - \lambda (\Phi^\dagger \Phi)^2. 
\nonumber 
\ee
$P_f({\cal D}_\Lambda^2)$ and $P_s({\cal D}_\Lambda^2)$ are again polynomial free of real ghosts. 
To achieve full finiteness of all running coupling constants we need few other local operators, which the interested reader can find in the references \cite{LeslawModestoFinite, PivaFGT}.
In contrast to the Lee-Wick standard model of particle physics previously proposed \cite{Wise1, Wise2}) where real ghosts move out the real axis at quantum level, here the complex conjugate poles are a feature of the classical theory. 
Moreover, the theory proposed in this section is super-renormalizable or finite at quantum level.

\section*{Conclusions and remarks}

The class of local higher derivative gravitational theories studeid in this paper have extra 
complex conjugate poles besides the standard massless graviton pole in the propagator. 
%Lee and Wick, and Cutkosky et al. provide a prescription for handling this issue. 
Theories with complex conjugate poles in the propagator seem to be well defined as shown in 
%Lee and Wick, and Cutkosky et al. 
\cite{LW, Cutk, Jansen, Yamamoto1, Veltman} especially if super-renormalizable or finite. 
The extra unphysical particles associated with the new poles are not in the physical Hilbert space for the
asymptotic states, but are forced to decay in ordinary gravitational degrees of freedom by the real energy
conservation.

At quantum level the higher derivative operators make the theory super-renormalizable in any dimension. 
Indeed only one-loop up to three-loops divergences could be present depending on the particular set of higher derivative operators included in the action. 
However, for the case of a one-loop super-renormlizable theory, 
a local potential starting cubic in the Riemann tensor 
does not affect the propagator around the flat spacetime, but makes all the beta functions to 
vanish, and the theory turns out to be finite.  
Moreover, % the tensorial structure of pure gravity enables us to conclude that
using dimensional regularization the theory is finite in odd dimension 
%as a matter of fact 
because there are no local one-loop counterterms with an odd number of derivatives in odd dimension.  
%The action in odd dimension is very simple because we can consistently fix to zero many of the coupling
We here again show a $D$-dimensional minimal prototype theory,
\be
 %\hspace{-0.5cm}
\mathcal{L} =  - 2 \kappa_D^{-2} \sqrt{|g|} \Big[ %\frac{2}{\kappa_D^{2}} %\sqrt{|g|} 
%\Big( 
R - s_0 \, 
G_{\mu \nu}  \Box^{\gamma+ \frac{D}{2} - 2 } %}{\Lambda^2}
 R^{\mu \nu} +  
\sum_{i=1}^{n_K} s_i \,  {\mathcal R}^\frac{D}{2} \, \nabla^{2 \gamma -4} \, {\mathcal R}^2 \Big]  %\Big) 
\, , 
\label{finale}
 \ee 
where $s_0 = %(-1)^{\gamma+{\rm N}}
(-1)^{\gamma + (D-4)/2}/\Lambda^{2\gamma+D-2}$, and the sum is over the minimal number ``$n_K$" of killer operators we need to make the theory finite.
Within the quantum field theory framework this theory preserves Lorentz and diffeomorphism invariance, and respect Lee-Wick unitarity 
in the subspace of real physical states. Furthermore, (\ref{finale}) is finite in odd dimension and 
super-renormalizable in even dimension for any choice
of the parameters $s_0, s_i$. 
Moreover, for particular choice of the parameters $s_i$ all the beta 
functions can be make to vanish in $D=4$ and likely in any even dimension.
Therefore the theory turns out to be finite.

%Using the background field method, the Barvinsky-Vilkovisky techniques, and in the Batalin-Vilkovisky formalism %it is straightforward to prove 
%we can infer about unitarity for the tree-level quantum effective action at any perturbative order in the
%loop expansion. 
In our theory tree-level unitarity is guaranty by the real energy conservation that comes together with the S-matrix definition. 
In other words the complex conjugate poles never go on shell and the optical theorem is satisfied 
on the real physical and Lorentz invariant subspace. At quantum level the CLOP prescription guarantees 
unitarity at least for the minimal super-renormalizable theory \cite{Cutk, Wise3}.

 %A truncation of the classical equation of motion has been studied in a company paper
%showing regularity of the spacetime at short scale with a replacing of the black hole singularity 
%with a deSitter core for whatever the higher derivative polynomial theory is. 

The singularities that plague the gravitational potential of Einstein gravity
are here smeared out because of the soft behaviour of the propagator at short distance.
The complex conjugate particles contribute to overall cancel out the divergent contribution %at short distance due to 
of the massless physical graviton field \cite{tiberio}.

%

%\section{Remarks}

%In this paper we dealt with purely gravitational theories, therefore 
%following the van Tonder \cite{vanTonder} argument it comes natural to interpret 
%the complex conjugate poles as ghost black holes. 
%From two graviton scattering amplitude it is created a ghost black hole pair 
%that in turn decays into two gravitons again without ever appear on-shell. %in the asymptotic states. 
%unitarity violation in Lee-Wick quantum field theory.
%{\color{red} Since we have microcausality violation the particles seen by the future observer are emitted from a point causally prior to the collision of the incoming matter. This is reminiscent of the way Hawking radiation originates causally prior to the singularity that absorbs the incoming matter in a black hole. }

%In loops the energy runs to infinity and 
{Let us further expand on the interpretation of complex conjugate poles. % this sharp interpretation. 
%We all experience the gravitational force in the everyday life because of the 
%anti-screening effect of gravitons (gravitons warp the spacetime itself.) 
In the theory here proposed, by increasing the energy, gravity becomes stronger, but %However, 
in the short distance 
limit $\ell <<1/\Lambda \sim 1/M_P$%week gravity in the everyday life, but in the high energy regime a gravitational anti-screening   
%effect is at work making gravity stronger and the spacetime curvature large.
%
%At low energy gravity is week, but at 
%At high energya graviton anti screening effect is at work making gravity stronger and the spacetime curvature bigger. 
%
%In the 
%limit of ${\rm energy} >> \Lambda \sim M_P$
\footnote{We have two scales in our theory $G_N$ and the length 
scale $\ell = 1/\Lambda$, but here they are identified in order to simplify the discussion.}
gravity becomes weak again (constant gravitational potential and zero gravitational force) due to the 
anti-screening effect of the gravitons, which wins over 
the screening effect of the virtual anti-graviton particle pairs. 
In other words, getting closer to the mass the anti-screening effect of the surrounding gravitons diminishes, so the 
full contribution of this effect would be increasingly weak and 
%be to make weaken 
the ``effective mass" will decrease with decreasing distance.
%Since the gravitons increase the effective mass when we go very close to it 
This is analog to what occurs in quantum cromo-dynamics where the quarks play the role of anti-gravitons and
gluons the role of gravitons.
%Finally the screening and anti-screaming effects delete each other and gravity disappears, 
%as evident from the coincident limit in the two point function. 
}% in coordinate space.  
Similarly, also the attractive gravitational force increases with the energy, but vanishes in the
zero separation distance limit because of the repulsion due to the anti-gravitons. 
In the intermedium energy regime such unstable unphysical %particle 
pairs, the anti-gravitons, are excited without to go on shell. 
This is reminiscent of classical radiation surrounded by a 
complete absorber. In a complete absorber, radiation has to be absorbed and no asymptotic photons exist. 
Therefore, a quantum theory with complex conjugate poles would not have the associated 
asymptotic particles, as a simple calculation of the absorptive part shows. 
%In our theory the gravitons themselves give rise to an high curvature bubble absorber 
%in which anti-gravitons are confined. 

{On the footprint of the Calmet's proposal 
\cite{Calmet1, Calmet2, Calmet3, Calmet4, 'tHooft:1992zk} we can give here  the following 
alternative (or maybe equivalent) interpretation to the complex conjugate poles: they actually are the mass and the width of light black holes precursors, 
\be
k_0^2 = \left( M_{\rm BH} - i \frac{\Gamma_{\rm BH}}{2} \right)^2 \, .  
\ee 
In our example (\ref{PY}) $\eta^2$ is identified with the above pole, %M_{\rm BH}=\cos \pi/8
while the complex conjugate leads to the acausal effects.
In other words, our local theory describes the usual massless graviton and a finite number of micro black holes.
This idea can be supported evaluating the classical equations of motion. 
Indeed, it has been shown in \cite{yaudong} that the Schwarzschild and Kerr black holes are exact solutions of the theory. 
Following the t'Hooft suggestion we could say that the theory here proposed is a kind of ``unitarization" of the higher derivative Stelle's gravity through out the explicit introduction of virtual black holes (see below the comparison with the Stelle's quadratic gravity at one loop.)

In a nonlocal super-renormalizable theory we expect the same structure not at the classical
level, but for the quantum action (for example, this can be read out of the finite $log$ contributions to the quantum action \cite{ComplexGhostsIS}.)
However, here the number of complex conjugates poles is infinite allowing for a spectrum of 
arbitrary large black holes.} %This poles are actually Landau poles probably outside the perturbative regime. 

In Stelle theory we have the same phenomenon because the real ghost pole splits into two complex conjugate poles at quantum level. Again we can give the same interpretation and infer that at one-loop the spectrum of the quantum action is compatible with unitarity and the real ghost is converted in a pair of particles consisting on a black hole and the complex conjugate state 
\cite{Tomboulis:1977jk, Tomboulis:1980bs, Smolin:1981rm}. 
% micro-causality is violated while macro-causality is preserved. 
% and the real ghost is converted in a pair of black holes. 

Let us notice that the above interpretation is based on a one-loop computation, therefore it is only perturbative, 
and in Stelle theory we need to compute higher loop corrections to show the stability of the spectrum. 
However, in a super-renormalizable theory (convergent for $L>1$) the beta functions are one-loop exact 
%since the theory is 
and the asymptotic freedom makes the interpretation likely correct at any perturbative order. 
% free it should be correct at any perturbative order. 
Indeed, finite perturbative contributions to the quantum action can only slightly move the complex conjugates poles.

%The perfect absorber has here characteristic size $1/\Lambda \sim 1/M_P \sim \ell_P$, so 
%it comes natural to read it as the event horizon for a black hole created through the high energy gravitons
%scattering. However, the anti-gravitons prevent the formation or the stability of such non-perturbative
%black hole system and it is forced to decay or ``evaporate" through the emission of gravitons. 
 % In other words, 
  Following the van Tonder \cite{vanTonder} argument, or suggestion, it comes natural (by CPT invariance of the theory) to interpret a full scattering process as the creation and evaporation of a black hole system. 
%the anti-gravitons as virtual black holes. 
When two gravitons (or matter particles) scatter, a black hole is created, together with a 
%n anti-black hole (
white hole (the CPT conjugate solution), 
that in turn decays into two gravitons again (or matter particles) without ever appear on-shell.

%in the asymptotic states.
%unitarity violation in Lee-Wick quantum field theory.
Since we have microcausality violation or more properly effective non-locality in time \cite{vanTonder}, the particles seen by the future observer are emitted from a point causally prior the collision of the incoming matter
(which happens at the singularity.) This is indeed reminiscent of the way Hawking radiation originates causally prior the singularity that absorbs the incoming matter in the black hole. The scattering process happens at the singularity, while the 
outgoing particles are earlier generated at the event horizon. 
On the other hand, the particles are emitted in the region near the singularity before the 
Hawking scattering process (CPT revers of the Hawking emission) can accur at the horizon. 

The spacetime structure is obtained replacing the event horizon with a simply connected trapped surface
achieved gluing together the black hole and white hole horizons. 

%By CPT invariance, it comes natural to interpret the complex conjugate (run-away inclreasing) pole as a white hole which is crated 

This interpretation of complex conjugate pairs as describing black holes - white hole pairs 
seems compatible with %remind us 
the t'Hooft complementary principle \cite{'tHooft:1984re} as %a because it is 
a consequence of the
CPT invariance of the theory. 

%We can support this interpretation with an explicit computation. It is common feature of the theories 
%here proposed to develop linear real ghost instabilities around approximate black hole solutions.
%However, this is not a catastrophic issue because the higher derivative (or in sense nonlocal) nature of the theory.
%Indeed the  

Finally, we hope that the local action here proposed will stimulate cosmologists and people of the black hole community in starting looking for exact solutions and eventually infer about their stability. 
The minimal theory here proposed  
is ``just six order" in derivatives of the metric, therefore a classic study of the theory can be fielded
relatively easily. 
%In particular the stability analysis of the exact classical solutions will shed on the final true about complex conjugate poles. 

\section*{Acknowledgments}
A special thanks goes to Ilya L. Shapiro for drawing our attention to higher derivative theories 
involving complex conjugate poles. 
%a large body of literature on 
%nonlocal field theory. 
We also thank Damiano Anselmi and Leslaw Rachwal for asking numerous stimulating questions. 
% and stimulating my commitment stimulate my efforts to understand more deeply

\end{document}